\documentclass[aps,prb,superscriptaddress,twocolumn,showpacs]{revtex4-1}
\usepackage{graphicx,amsmath,hyperref}
\usepackage[version=3]{mhchem}

\begin{document}

\title{Photoionization cross section of {\emph 1s} orthoexcitons in cuprous oxide}

\author{Laszlo Frazer}
\email{prb@laszlofrazer.com}
\affiliation{Department of Physics, Northwestern University, 2145 Sheridan Road, Evanston, Illinois 60208-3112, USA.}

\author{Kelvin B. Chang}
\affiliation{Department of Chemistry, Northwestern University, 2145 Sheridan Road, Evanston, Illinois 60208-3112, USA.}

\author{Kenneth R. Poeppelmeier}
\affiliation{Department of Chemistry, Northwestern University, 2145 Sheridan Road, Evanston, Illinois 60208-3112, USA.}
\affiliation{Chemical Sciences and Engineering Division, Argonne National Laboratory, 9700 South Cass Avenue, Argonne, Illinois 60439, USA}
\author{John B. Ketterson}
\affiliation{Department of Physics, Northwestern University, 2145 Sheridan Road, Evanston, Illinois 60208-3112, USA.}
\affiliation{Department of Electrical Engineering and Computer Science, Northwestern University, 2145 Sheridan Road, Evanston, Illinois 60208-3112, USA.}
\date{\today}

\begin{abstract}
	We report measurements of the attenuation of a beam of orthoexciton-polaritons by a photoionizing optical probe.  Excitons were prepared in a narrow resonance by two photon absorption of a 1.016 eV, 54 ps pulsed light source in cuprous oxide (\ce{Cu2O}) at 1.4 K.  A collinear, 1.165 eV, 54 ps probe delayed by 119 ps was used to measure the photoionization cross section of the excitons.   Two photon absorption is quadratic with respect to the intensity of the pump and leads to polariton formation.  Ionization is linear with respect to the intensity of the probe.  Subsequent carrier recombination is quadratic with respect to the intenisty of the probe, and is distinguished because it shifts the exciton momentum away from the polariton anticrossing; the photoionizing probe leads to a rise in phonon-linked luminescence in addition to the attenuation of polaritons.  The evolution of the exciton density was determined by variably delaying the probe pulse.  Using the probe irradiance and the reduction in the transmitted polariton light, a cross section of $(3.9\pm0.2)\times 10^{-22}$ m$^2$ was deduced for the probe frequency.
\end{abstract}


\maketitle

\section{Introduction}

Ionization can be used to investigate exciton structure and dynamics.  Photoionization also has possible importance as a probe of exciton Bose Einstein condensates \cite{ideguchi2008coherent} prepared by two photon absorption and as a probe of the exciton density.\cite{yoshioka2010quantum}  The exciton photoionization cross section can be used to design exciton-exciton scattering and two photon absorption experiments which are free of interference from photoionization.

A cuprous oxide yellow \emph{1s} orthoexciton is an excellent quantum system for studies of photoionization since: 1) it shares a Rydberg-like structure with hydrogen \cite{hall1936theory} and positronium;\cite{fee1991sensitive} 2) it is relatively stable to radiative decay;\cite{mysyrowicz1979long}  3) there is extensive interest in strongly interacting conditions;\cite{laszlo2013unexpectedly,yoshioka2011transition,o1999auger,roslyak2012coherent,naka2013free} 4) nonlinear optical methods can be used to prepare the excitons;\cite{mani2010nonlinear,goto1997bose,ideguchi2008coherent} and 5) the polariton (a mixed exciton/photon) state can be studied selectively.\cite{frohlich1991coherent} Exciton photoionization was inferred from photoconductivity in the organic semiconductors anthracene,\cite{courtens1967photo,kepler1967photoionization} tetracene,\cite{schlotter1977photoionization} and $p$-terphenyl,\cite{morikawa1983generation}  although the electronic structures of these materials differ significantly from that of cuprous oxide.

In the experiments reported here photon pairs from a 1.016 eV pump pulse were combined through a narrow resonance in a cuprous oxide single crystal to produce propagating 2.0335 eV (609.71 nm) orthoexciton-polaritons. Preparing the excitons via two photon absorption ensures they are in a well defined, non-thermal state.  Conservation laws forbid exciton decay outside the polariton regime.

Previous studies of two photon absorption in cuprous oxide hinted at the existence of a photoionization process.  The process was masked, however, by: 1) third harmonic generation,\cite{ideguchi2008coherent,frazer2014third} 2) Auger exciton-exciton annihilation,\cite{laszlo2013unexpectedly} and 3) identical pump and probe photons.

\begin{figure}
	\includegraphics{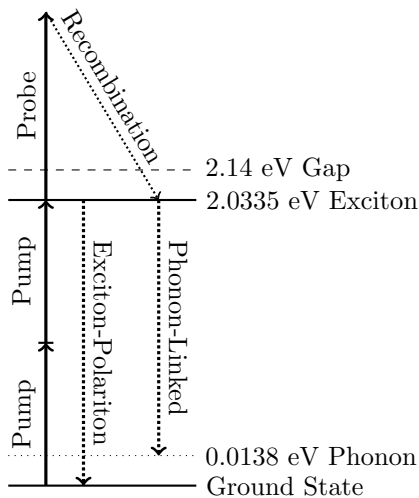}
	\caption{Energy level diagram illustrating two photon absorption of the pump, exciton ionization by the probe, nonradiative electron/hole recombination, phonon-linked luminescence, and exciton-polariton sum frequency.}
	\label{fig:level}
\end{figure}

To distinguish between these processes we propagate, in addition to the 1.016 eV pump pulse, a temporally and energetically distinct 1.165 eV probe pulse (Fig. \ref{fig:level}) along the same crystal axis. The purpose of the probe pulse is to ionize excitons generated by the pump pulse. This contrasts with Terahertz spectroscopy of Lyman transitions,\cite{jorger2003infrared,kubouchi2005study,kuwata2004time,tayagaki2005yellow,ideguchi2008coherent} where the electron and hole typically remain bound.  The probe pulse is not directly absorbed in the absence of the pump.

We measure the time averaged, transmitted quadrupole exciton-polariton population, which only samples those excitons permitted to decay by the selection rules.\cite{yoshioka2006dark,elliott1961symmetry} The probe beam strongly attenuates the quadrupole orthoexciton-polaritons.  Furthermore, this attenuation depends on the temporal positioning of the probe pulse.  Auger recombination is excluded through the use of a probe photon which does not produce excitons.  $\chi^{(3)}$ processes are also ruled out when the pump and probe are temporally separated.

Inelastic scattering processes are always accompanied by elastic scattering.\cite{landau1977quantum} In addition to ionization, inelastic processes can include various final states, such as excited excitons.\cite{kuwata2004time,schmutzler2013signatures} To our knowledge the cross section for photon-polariton scattering has not been studied theoretically although polariton-polariton scattering has been considered.\cite{combescot2007polariton} As a simple model, we regard the exciton as a hydrogenic atom for which the ionization cross section of the \emph{1s} state is given by \cite{hall1936theory}
\begin{align}
	\sigma&=\frac{ha_0}{\mu c}\frac{2^9\pi}{3}\frac{e^{-4\frac{v_0}{v}\arctan\left( \frac{v}{v_0} \right)}}{\left( 1+\frac{v^2}{v_0^2} \right)^4\left( 1-e^{-2\pi\frac{v}{v_0}} \right)}
\end{align}
where 
\begin{align}
	\frac{v^2}{v_0^2}&=\frac{h\nu}{\operatorname{Ry}}-1
\end{align}
Here $h$ is Planck's constant, $a_0= $ 0.79 to 1.11 nm is the exciton Bohr radius,\cite{tayagaki2005yellow} $\mu = 4\times 10^{-31}$ kg is the reduced effective mass,\cite{moskalenko2000bose} $c$ is the speed of light, $h\nu$ is the photon energy, and Ry$=97$ meV is the exciton Rydberg energy.\cite{moskalenko2000bose} At the probe photon energy the measured cross section is $(3.9\pm0.2)\times 10^{-22}$ m$^2$, which is larger than the range for the hydrogenic atom model of 1.3 to $2.7\times10^{-23}$ m$^2$ (depending on the value of the Bohr radius) and, in addition, about 7.5 times larger than the reported values for $p$-terphenyl \cite{morikawa1983generation} and anthracene.\cite{kepler1967photoionization} The cross section is only $(5\pm 2)\times10^{-26}$ m$^2$ in tetracene.\cite{schlotter1977photoionization} The cross section determined here only applies to excitons in the population sampled through polariton coupling, which is highly collimated in the forward direction; it includes all processes caused by the probe beam which diminish the size of that population.

\section{Methods}
\subsection{Pump/Probe Scheme}
\begin{figure*}
	\includegraphics[width=.9\textwidth]{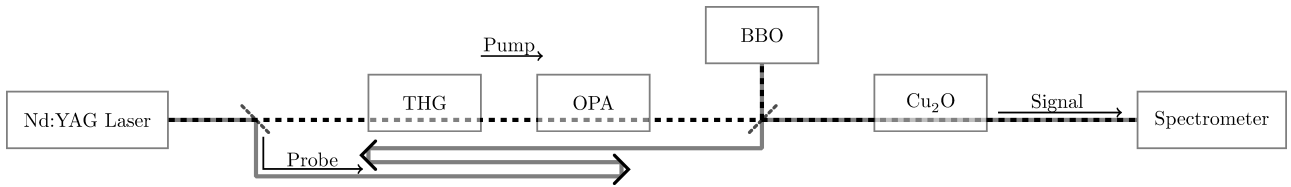}
	\caption{Experiment diagram.  The laser beam is split internally.  The branch shown at the top is converted to third harmonic by cascade sum frequency generation (THG) and then sent to an optical parametric amplifier (OPA).  The OPA idler beam is tuned to half the \emph{1s} orthoexciton energy.  The other branch is delayed.  The beam branches recombine, with half passing through the cuprous oxide sample and the other half applied to a $\beta$-Barium Borate (BBO) crystal or a power meter in the reference arm.}
	\label{fig:diagram}
\end{figure*}

Fig. \ref{fig:diagram} is a diagram of the apparatus.  The frequency-tripled output of a mode-locked Nd:YAG laser is used to pump an optical parametric amplifier (OPA). The full width half maximum (FWHM) pulse duration is $(5.4\pm 0.5)\times 10^{-11}$ s, which is determined by sum-frequency cross-correlation of the pump and probe in $\beta$-barium borate.   It is assumed to be the same for both pulses.  The repetition rate is 10 Hz. 

At the two-photon resonance energy of 1.016 eV (1220 nm), the spectral bandwidth of the linearly polarized idler output of the OPA has a FWHM of 9.28 meV, far broader than the line width of the orthoexciton-polariton.\cite{kuwata2002phase,ideguchi2008coherent,yoshioka2014selective} The two photon absorption of this laser was previously characterized.\cite{mani2010nonlinear} We note that two photon sum-frequency generation in cuprous oxide (which is centrosymmetric) operates differently from the mechanism of second harmonic generation in noncentrosymmetric materials.\cite{halasyamani1998noncentrosymmetric}  Sum frequency generation is permitted for the case of electric quadrupole and magnetic dipole active states,\cite{epperlein1987second} the orthoexciton polariton corresponding to the former.\cite{frohlich2005high}  

The 1.165 eV (1064 nm) probe beam is obtained directly from the Nd:YAG laser and, after passing through a delay line, is joined with the pump beam in a 50:50 nonpolarizing beam combiner, with half the combined beam proceeding to the sample with normal incidence. 

\subsection{Samples}

High quality samples are required to produce polariton sum-frequency generation.  For this work, a high purity synthetic crystal was grown using the floating zone method.\cite{chang2013removal}  Copper vacancies and cupric oxide inclusions were further minimized by annealing.  The primary sample was grown from 0.99999 purity Cu in air, at 4.5 mm/hour with 7 RPM counterrotation.  It was cut with [111] faces and annealed at 1045 $^\circ$C for 5 days with a 5 $^\circ$C/minute ramp rate.  The final sample thickness after polishing was 3.988 mm.  The data shown in Fig. \ref{fig:100idler} are from an additional sample with [100] faces that was grown at 3.5 mm/hour and annealed for only 3 days.  The thickness was 0.24 mm.  This sample was previously reported in Ref. \onlinecite{chang2013removal}, which describes the details of the growth process.  Note that polarization selection rules apply to [100] pump propagation and that the greater thickness of the [111] sample leads to stronger two photon absorption.  Neither sample is perfectly phase pure.

\subsection{Measurement}
A near-field 0.52 mm radius aperture was used to select the center of the beam so that the sample was illuminated approximately uniformly, which also assists in the alignment.  The sample was submerged in superfluid helium at $1.4\pm 0.5$ K. The orthoexciton-polariton sum frequency beam was measured in transmission and averaged across laser shots with an Andor 303 mm focal length Czerny-Turner spectrograph and DU420A-BEX2-DD thermoelectric cooled CCD camera.  The second half of the combined beam was monitored, after a matched aperture, using a power meter or a $\beta$-barium borate sum frequency generator, slightly phase mismatched for each of the three sum frequencies. A spectrometer was used to measure the sum frequencies.

\section{Results}

\begin{figure}
	\includegraphics[width=\columnwidth]{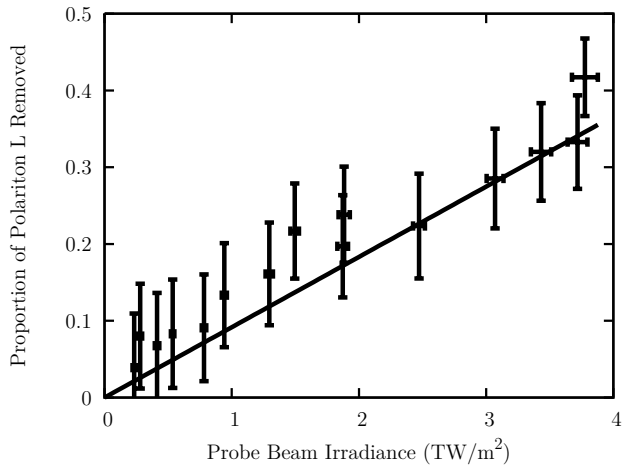}
	\caption{Exciton photoionization.  The proportion of the transmitted polaritons removed increases linearly with the number of incident probe photons.}
	\label{fig:power}
\end{figure}

The presence of the probe beam resulted in a substantial reduction in the observed exciton-polariton sum frequency (Fig. \ref{fig:power}).  
\subsection{Spectra}
\begin{figure}
	\includegraphics[width=\columnwidth]{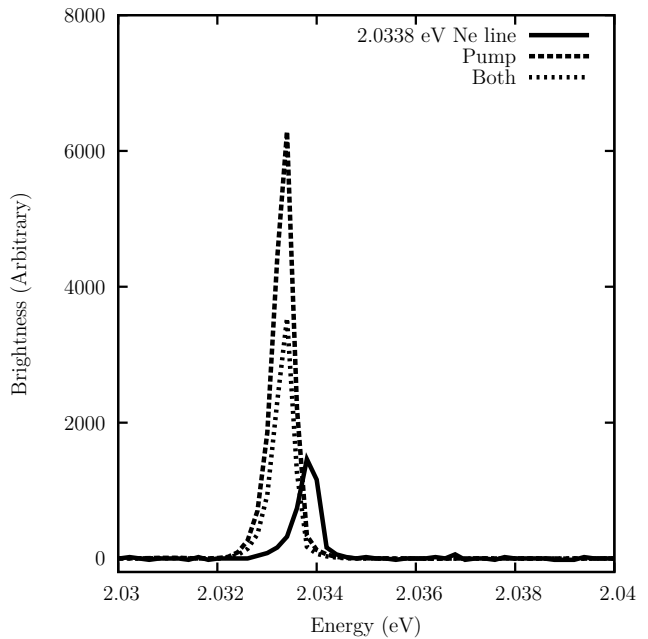}
	\caption{Exciton-polariton spectrum, with pump beam only and with both pump and probe beam.  A spectral line from a neon lamp is used to confirm calibration.\cite{NIST_ASD}  The neon brightness is not comparable. The line widths are resolution limited.}
	\label{fig:spectrum}
\end{figure}

Based on a Gaussian model, the polariton peak in the spectrum was at 2.0335$\pm$ 0.0002 eV for a lattice temperature of $1.4\pm 0.5$ K, independent of probe beam scattering (Figure \ref{fig:spectrum}).  The shape of the exciton-polariton spectrum, which is instrument resolution limited, showed no change during probing.   Temperature dependence has been investigated.\cite{ito1997detailed}  The line widths were resolution limited.  Measurements were typically collected in a lower resolution mode, which improved statistics.  Light was detected more efficiently by opening the spectrometer entrance slit, thereby increasing the sensitivity of the instrument.  The resulting decrease in the instrument resolution did not allow for the exciton temperature to be measured by fitting the phonon-linked luminescence spectrum.  In this case the polariton brightness $L$ was determined by integrating from 2.029 eV to 2.038 eV.  The phonon-linked luminescence was integrated from 2.016 to 2.026 eV.  

\begin{figure}
        \includegraphics[width=\columnwidth]{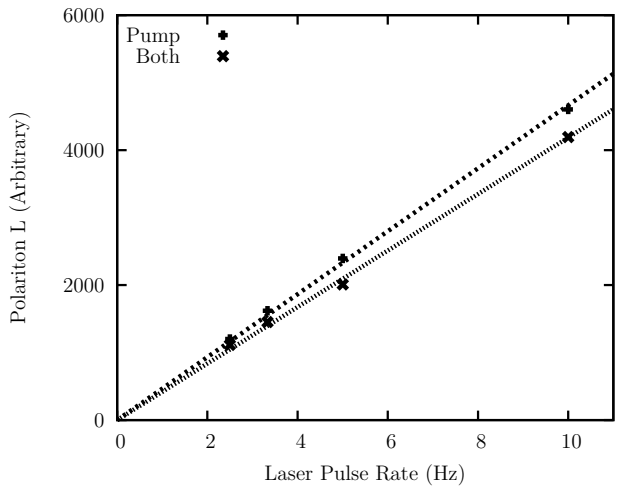}
        \caption{Polariton intensity vs. laser pulse rate showing that the signal increases linearly.}
        \label{fig:pulserate}
\end{figure}

An average heating of the sample by the laser was ruled out as a cause of the reduction in polaritons detected by varying the laser pulse rate.  No change in the scattering cross section was observed (Figure \ref{fig:pulserate}).  If the laser pulses resulted in an average heating of the sample, the intensity of the signal would fall below the expected linear dependence at high pulse rates.

\subsection{Experimental Cross Section Calculation}
\begin{figure}
	\includegraphics[width=\columnwidth]{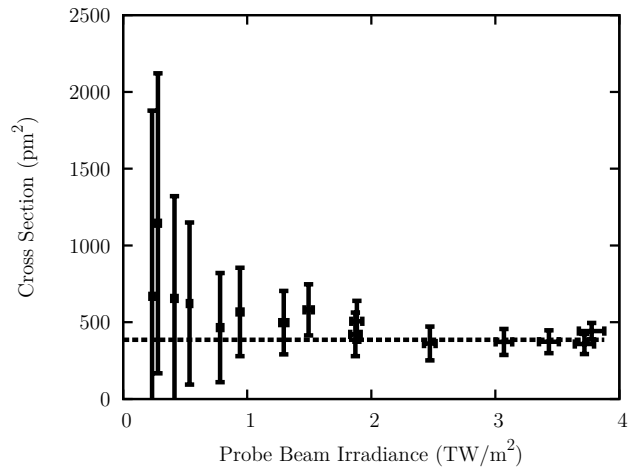}
	\caption{Photoionization cross section values associated with Fig. \ref{fig:power}.  The average value is marked with a dashed line.}
	\label{fig:crosssection}
\end{figure}

As the number of probe photons in each pulse increased, the proportion of the polaritons that were removed increased linearly (Fig. \ref{fig:power}); here $N_i$ is the number of incident photons in the probe laser pulse and $L$ is the wavelength-integrated exciton-polariton sum frequency intensity with the probe beam on or off.  For a working area $\mathcal{A}$, the cross section (Fig. \ref{fig:crosssection}) is 
\begin{align}
	\sigma&=\frac{\mathcal{A}}{N_i}\left( 1-\frac{L_{\text{on}}}{L_{\text{off}}} \right)&=(3.9\pm0.2)\times 10^{-22} \text{ m}^2
\end{align}
The pump energy for the data shown in Figs. \ref{fig:power}, \ref{fig:crosssection} and \ref{fig:phonon} is 23.8 $\mu$J per pulse.  The probe pulse is delayed $119\pm1$ ps to ensure pump absorption is complete and spontaneous exciton decay is minimal. Since the probe pulse arrives after the pump pulse, there is no significant error due to temporal overlap of the pulses.  Systematic error due to emission detected before the probe pulse was kept small. Short term stabilities for the pump and probe beams were 95\% and 98\% respectively.  

\subsection{Phonon-Linked Luminescence}
\begin{figure}
	\includegraphics[width=\columnwidth]{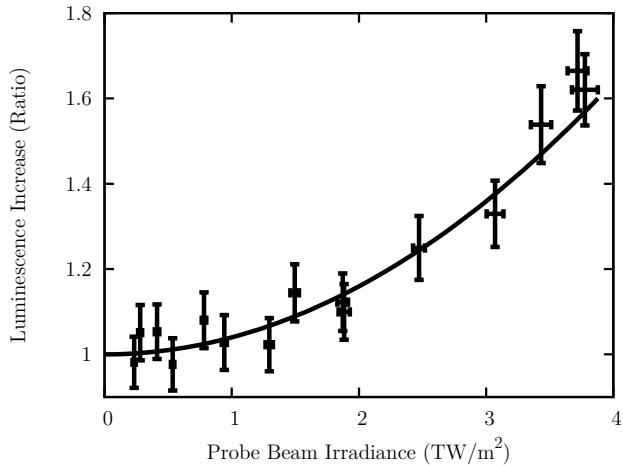}
	\caption{Phonon-linked luminescence.  The luminescence increases quadratically (curve) with the number of incident probe photons. This quadratic behavior is a result of the post-ionization decay chain of two excitons.}
	\label{fig:phonon}
\end{figure}

The carriers produced via photoionization recombine to produce phonon-linked luminescence, which increased quadratically with the probe beam intensity (Fig. \ref{fig:phonon}), because the binding rate is proportional to the product of the carrier densities.  This is evidence of a double photoionization assisted two exciton decay, which requires a total of six absorbed photons.  In Fig. \ref{fig:phonon}, the phonon-linked luminescence is reported as a ratio of the measurement with both the pump and the probe to a measurement with the pump alone.

Third harmonic generation \cite{frazer2014third} from the probe beam can also produce phonon-linked luminescence.  However, no three photon sum frequency generation in the sample was observed by direct detection.  

Phonon-linked luminescence is not directed in a beam because recombined excitons have random momentum.  Therefore the phonon-linked luminescence was much weaker than the (directed) exciton-polariton signal.  Excitons which cannot contribute to the polariton signal, such as the paraexciton state, may contribute to the quadratic increase in phonon-linked luminescence as they are ionized.

\subsection{Dynamics}
\begin{figure}
	\includegraphics[width=\columnwidth]{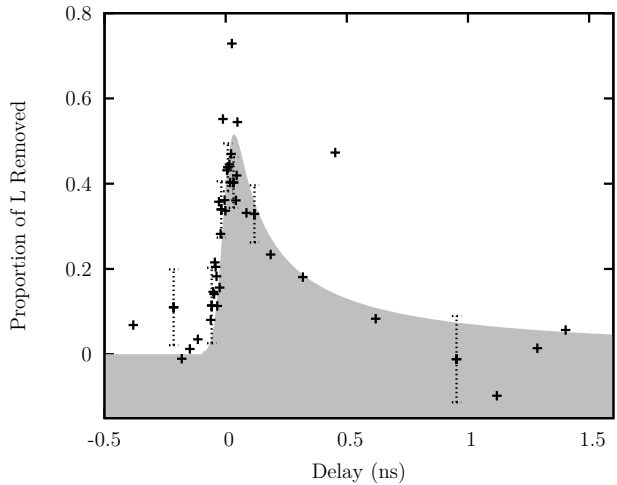}
	\caption{Probe pulse delay.  The amount of exciton ionization at the probe pulse arrival time indicates the relative exciton density at that time.  A sample of error bars is shown for clarity. The curve represents pumping with a Gaussian temporal profile followed by Auger decay.}
	\label{fig:delay}
\end{figure}

If the ionization pulse arrived before the pump pulse, no ionization occurred (Fig. \ref{fig:delay}).  If the ionization pulse arrived after the pump pulse, the apparent photoionization cross section decreased rapidly owing to dispersion \cite{frohlich1991coherent} (and the resulting spread in exciton group velocities) and Auger decay \cite{laszlo2013unexpectedly,o1999auger,yoshioka2010quantum} of the exciton-polaritons.  A density dependence indicating Auger decay was demonstrated previously under similar conditions.\cite{mani2010nonlinear}  The rise time is explained by the temporal structure of the laser pulses, not by a change in degenerate spin state.\cite{yoshioka2006dark}  The pump pulse was 25 $\mu$J with a long term standard deviation of 1 $\mu$J.  The probe pulse, which is subject to divergence owing to the additional path length used to generate the delay, was 116 $\mu$J with a long term standard deviation of 35 $\mu$J.

\subsection{Pumping}
The magnitude of the polariton signal depends on the spectrum of the photons in the pump beam, but the cross section was independent of the mean pump photon energy.  The state pumped by two photon absorption is presumed to be independent of the pump photon energy and spectral phase, provided there is absorption.\cite{yoshioka2014selective}  

\begin{figure}
        \includegraphics[width=\columnwidth]{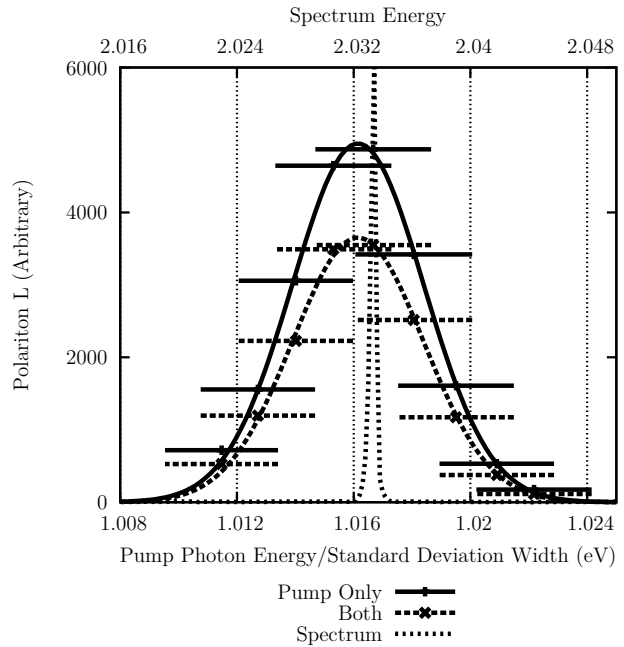}
        \caption{Exciton-polariton intensity detected as a function of two photon pumping energy with and without the probe beam.  Bars are the standard deviation pump beam spectral width, not the uncertainty in the mean energy.  }
        \label{fig:opa}
\end{figure}

The pump spectrum center was adjusted using the OPA (Figure \ref{fig:opa}).  No change in the cross section was found in the small region where there is absorption.  The bars indicate laser standard deviation line width, not uncertainty.  Pump spectra were determined by second harmonic generation using a $\beta$-barium borate crystal.  The upper curve is a Gaussian fit.  For the lower curve, which includes the probe beam, only the amplitude is fitted; the other parameters remain the same.  The peak pump photon energy lies below half the polariton energy, but the discrepancy can be explained by systematic uncertainties.  A resolution-limited polariton spectrum is shown.  In all other measurements reported here, the optimal pump photon energy is used.

\subsection{Selection Rules}
\begin{figure}
        \includegraphics[width=\columnwidth]{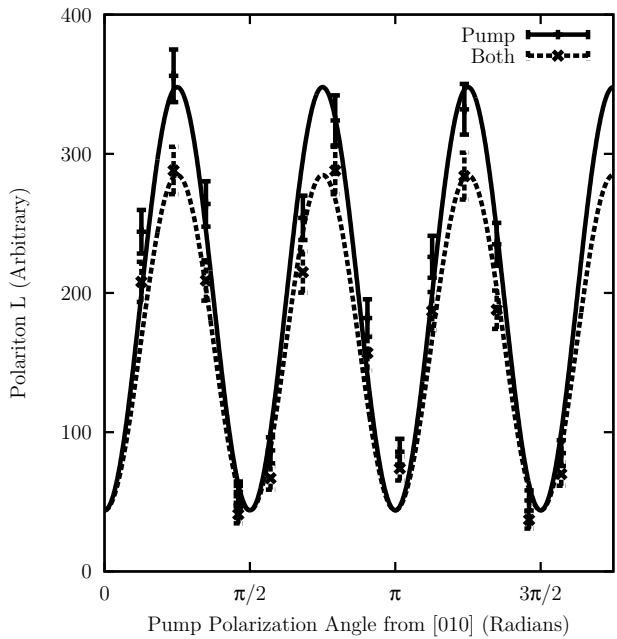}
	\caption{For [100] incidence, scattering versus the polarization of the pump photon.  The curves show the absorption selection rule \cite{yoshioka2006dark} with and without the probe pulse.  The difference between the data produced using only the pump beam and the data produced with both a pump and a probe beam is consistent with ionization of a fixed proportion of the excitons by the probe beam.}
        \label{fig:100idler}
\end{figure}

\begin{figure}
        \includegraphics[width=\columnwidth]{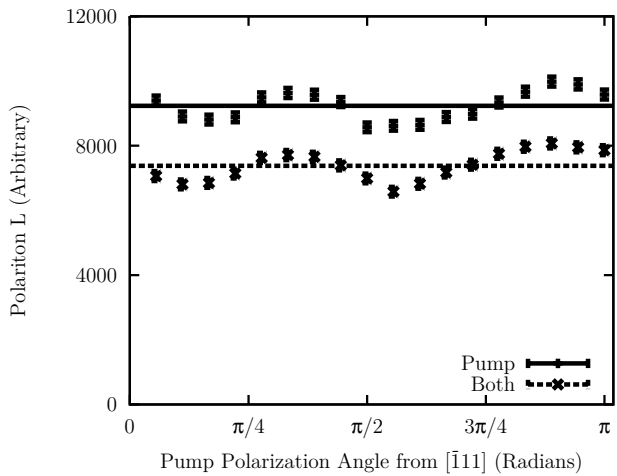}
        \caption{For [111] incidence, photoionization is essentially independent of the polarization of the pump photon.}
        \label{fig:pumppolar111}
\end{figure}

\begin{figure}
        \includegraphics[width=\columnwidth]{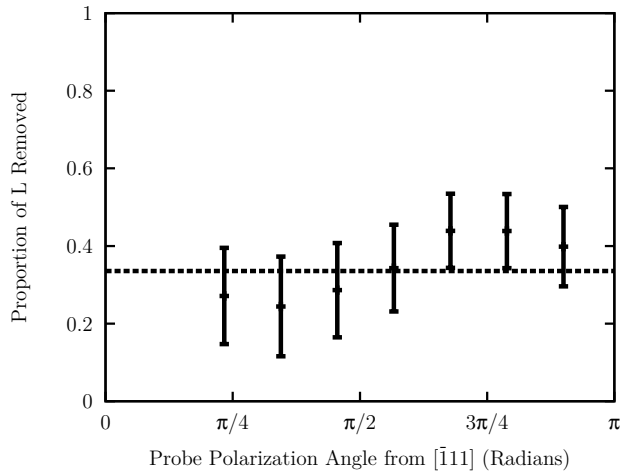}
        \caption{Proportion of the polariton signal removed versus probe polarization.  The measurements are consistent with a constant cross section.  This example is for [111] incidence.}
        \label{fig:ionizationpolar}
\end{figure}

The selection rules for two photon absorption have been previously investigated.\cite{yoshioka2006dark,elliott1961symmetry}  
The photoionization cross section did not vary in a statistically significant manner with the polarization of the pump or probe beams.
Figures \ref{fig:100idler} and \ref{fig:pumppolar111} show no significant change in the cross section as a function of pump laser polarization for [100] and [111] beam propagation respectively.  The curves in Fig. \ref{fig:100idler} do not quite reach zero as expected\cite{yoshioka2006dark,elliott1961symmetry} due to random error in the background subtraction or small systematic errors in crystal orientation and polarization.  Figure \ref{fig:ionizationpolar} shows no change in cross section with the polarization of the probe beam for [111] propagation. 


There has been extensive interest in cooling excitons to form a Bose-Einstein condensate.  Phase and polarization methods of reducing heating of the exciton gas owing to third harmonic generation have been developed.\cite{ideguchi2008coherent,frazer2014third}  The results suggest the photoionization process cannot be manipulated through polarization.  Phase manipulation is not expected to be useful because the Bohr radius of excitons is very short compared to the wavelength of photons used in typical experiments.  Photoionization heating of the exciton gas is best minimized by keeping the pump irradiance low or spatially separating it from the excitons being studied.  On the other hand, photoionization absorption will be a simpler, if destructive, alternative to Lyman spectroscopy for imaging exciton gases.\cite{jorger2003infrared,kubouchi2005study,kuwata2004time,tayagaki2005yellow,ideguchi2008coherent}

\section{Conclusions}

It is shown that \emph{1s} yellow orthoexciton-polaritons prepared by two photon absorption can be attenuated by probe photons.  This attenuation arises from ionization of the exciton component of the propagating polariton.  The resulting carriers produced luminescence through phonon emission.  By studying the relative temporal positioning of the pump and probe pulse, features associated with polariton dynamics \cite{frohlich1991coherent} are obtained.  Absorption of near infrared light in excited cuprous oxide might possibly be useful for energy capture in a normally transparent region of photovoltaic\cite{rakhshani1986preparation,wei2012photovoltaic,septina2011potentiostatic,lee2012growth} or photocatalytic devices.\cite{bendavid2013first,zhu2012cuprous,pal2013crystal,yoon2014manipulation}

\begin{acknowledgments}
We gratefully acknowledge NSF IGERT DGE-0801685.
This work was funded by the Institute for Sustainability and Energy at Northwestern (ISEN).
Crystal growth was supported by NSF DMR-1307698 and in part by Argonne National Laboratory under U.S. Department of Energy contract DE-AC02-06CH11357.
This work made use of the X-ray and OMM Facilities supported by the MRSEC program of the NSF (DMR-1121262) at the MRC of Northwestern.
K. C. was supported as part of the Center for Inverse Design, an Energy Frontier Research Center funded by the U.S. Department of Energy, Office of Science, Office of Basic Energy Sciences, under award number DE-AC36-08GO28308.
\end{acknowledgments}
\bibliography{photodis}

\begin{thebibliography}{44}%
\makeatletter
\providecommand \@ifxundefined [1]{%
 \@ifx{#1\undefined}
}%
\providecommand \@ifnum [1]{%
 \ifnum #1\expandafter \@firstoftwo
 \else \expandafter \@secondoftwo
 \fi
}%
\providecommand \@ifx [1]{%
 \ifx #1\expandafter \@firstoftwo
 \else \expandafter \@secondoftwo
 \fi
}%
\providecommand \natexlab [1]{#1}%
\providecommand \enquote  [1]{``#1''}%
\providecommand \bibnamefont  [1]{#1}%
\providecommand \bibfnamefont [1]{#1}%
\providecommand \citenamefont [1]{#1}%
\providecommand \href@noop [0]{\@secondoftwo}%
\providecommand \href [0]{\begingroup \@sanitize@url \@href}%
\providecommand \@href[1]{\@@startlink{#1}\@@href}%
\providecommand \@@href[1]{\endgroup#1\@@endlink}%
\providecommand \@sanitize@url [0]{\catcode `\\12\catcode `\$12\catcode
  `\&12\catcode `\#12\catcode `\^12\catcode `\_12\catcode `\%12\relax}%
\providecommand \@@startlink[1]{}%
\providecommand \@@endlink[0]{}%
\providecommand \url  [0]{\begingroup\@sanitize@url \@url }%
\providecommand \@url [1]{\endgroup\@href {#1}{\urlprefix }}%
\providecommand \urlprefix  [0]{URL }%
\providecommand \Eprint [0]{\href }%
\providecommand \doibase [0]{http://dx.doi.org/}%
\providecommand \selectlanguage [0]{\@gobble}%
\providecommand \bibinfo  [0]{\@secondoftwo}%
\providecommand \bibfield  [0]{\@secondoftwo}%
\providecommand \translation [1]{[#1]}%
\providecommand \BibitemOpen [0]{}%
\providecommand \bibitemStop [0]{}%
\providecommand \bibitemNoStop [0]{.\EOS\space}%
\providecommand \EOS [0]{\spacefactor3000\relax}%
\providecommand \BibitemShut  [1]{\csname bibitem#1\endcsname}%
\let\auto@bib@innerbib\@empty
\bibitem [{\citenamefont {Ideguchi}\ \emph {et~al.}(2008)\citenamefont
  {Ideguchi}, \citenamefont {Yoshioka}, \citenamefont {Mysyrowicz},\ and\
  \citenamefont {Kuwata-Gonokami}}]{ideguchi2008coherent}%
  \BibitemOpen
  \bibfield  {author} {\bibinfo {author} {\bibfnamefont {T.}~\bibnamefont
  {Ideguchi}}, \bibinfo {author} {\bibfnamefont {K.}~\bibnamefont {Yoshioka}},
  \bibinfo {author} {\bibfnamefont {A.}~\bibnamefont {Mysyrowicz}}, \ and\
  \bibinfo {author} {\bibfnamefont {M.}~\bibnamefont {Kuwata-Gonokami}},\
  }\href@noop {} {\bibfield  {journal} {\bibinfo  {journal} {Physical Review
  Letters}\ }\textbf {\bibinfo {volume} {100}},\ \bibinfo {pages} {233001}
  (\bibinfo {year} {2008})}\BibitemShut {NoStop}%
\bibitem [{\citenamefont {Yoshioka}\ \emph {et~al.}(2010)\citenamefont
  {Yoshioka}, \citenamefont {Ideguchi}, \citenamefont {Mysyrowicz},\ and\
  \citenamefont {Kuwata-Gonokami}}]{yoshioka2010quantum}%
  \BibitemOpen
  \bibfield  {author} {\bibinfo {author} {\bibfnamefont {K.}~\bibnamefont
  {Yoshioka}}, \bibinfo {author} {\bibfnamefont {T.}~\bibnamefont {Ideguchi}},
  \bibinfo {author} {\bibfnamefont {A.}~\bibnamefont {Mysyrowicz}}, \ and\
  \bibinfo {author} {\bibfnamefont {M.}~\bibnamefont {Kuwata-Gonokami}},\
  }\href@noop {} {\bibfield  {journal} {\bibinfo  {journal} {Physical Review
  B}\ }\textbf {\bibinfo {volume} {82}},\ \bibinfo {pages} {041201} (\bibinfo
  {year} {2010})}\BibitemShut {NoStop}%
\bibitem [{\citenamefont {Hall}(1936)}]{hall1936theory}%
  \BibitemOpen
  \bibfield  {author} {\bibinfo {author} {\bibfnamefont {H.}~\bibnamefont
  {Hall}},\ }\href@noop {} {\bibfield  {journal} {\bibinfo  {journal} {Reviews
  of Modern Physics}\ }\textbf {\bibinfo {volume} {8}},\ \bibinfo {pages} {358}
  (\bibinfo {year} {1936})}\BibitemShut {NoStop}%
\bibitem [{\citenamefont {Fee}\ \emph {et~al.}(1991)\citenamefont {Fee},
  \citenamefont {Mills~Jr}, \citenamefont {Shaw}, \citenamefont {Chichester},
  \citenamefont {Zuckerman}, \citenamefont {Chu},\ and\ \citenamefont
  {Danzmann}}]{fee1991sensitive}%
  \BibitemOpen
  \bibfield  {author} {\bibinfo {author} {\bibfnamefont {M.}~\bibnamefont
  {Fee}}, \bibinfo {author} {\bibfnamefont {A.}~\bibnamefont {Mills~Jr}},
  \bibinfo {author} {\bibfnamefont {E.}~\bibnamefont {Shaw}}, \bibinfo {author}
  {\bibfnamefont {R.}~\bibnamefont {Chichester}}, \bibinfo {author}
  {\bibfnamefont {D.}~\bibnamefont {Zuckerman}}, \bibinfo {author}
  {\bibfnamefont {S.}~\bibnamefont {Chu}}, \ and\ \bibinfo {author}
  {\bibfnamefont {K.}~\bibnamefont {Danzmann}},\ }\href@noop {} {\bibfield
  {journal} {\bibinfo  {journal} {Physical Review A}\ }\textbf {\bibinfo
  {volume} {44}},\ \bibinfo {pages} {R5} (\bibinfo {year} {1991})}\BibitemShut
  {NoStop}%
\bibitem [{\citenamefont {Mysyrowicz}\ \emph {et~al.}(1979)\citenamefont
  {Mysyrowicz}, \citenamefont {Hulin},\ and\ \citenamefont
  {Antonetti}}]{mysyrowicz1979long}%
  \BibitemOpen
  \bibfield  {author} {\bibinfo {author} {\bibfnamefont {A.}~\bibnamefont
  {Mysyrowicz}}, \bibinfo {author} {\bibfnamefont {D.}~\bibnamefont {Hulin}}, \
  and\ \bibinfo {author} {\bibfnamefont {A.}~\bibnamefont {Antonetti}},\
  }\href@noop {} {\bibfield  {journal} {\bibinfo  {journal} {Physical Review
  Letters}\ }\textbf {\bibinfo {volume} {43}},\ \bibinfo {pages} {1123}
  (\bibinfo {year} {1979})}\BibitemShut {NoStop}%
\bibitem [{\citenamefont {Frazer}\ \emph {et~al.}(2013)\citenamefont {Frazer},
  \citenamefont {Schaller},\ and\ \citenamefont
  {Ketterson}}]{laszlo2013unexpectedly}%
  \BibitemOpen
  \bibfield  {author} {\bibinfo {author} {\bibfnamefont {L.}~\bibnamefont
  {Frazer}}, \bibinfo {author} {\bibfnamefont {R.~D.}\ \bibnamefont
  {Schaller}}, \ and\ \bibinfo {author} {\bibfnamefont {J.}~\bibnamefont
  {Ketterson}},\ }\href@noop {} {\bibfield  {journal} {\bibinfo  {journal}
  {Solid State Communications}\ }\textbf {\bibinfo {volume} {170}},\ \bibinfo
  {pages} {34} (\bibinfo {year} {2013})}\BibitemShut {NoStop}%
\bibitem [{\citenamefont {Yoshioka}\ \emph {et~al.}(2011)\citenamefont
  {Yoshioka}, \citenamefont {Chae},\ and\ \citenamefont
  {Kuwata-Gonokami}}]{yoshioka2011transition}%
  \BibitemOpen
  \bibfield  {author} {\bibinfo {author} {\bibfnamefont {K.}~\bibnamefont
  {Yoshioka}}, \bibinfo {author} {\bibfnamefont {E.}~\bibnamefont {Chae}}, \
  and\ \bibinfo {author} {\bibfnamefont {M.}~\bibnamefont {Kuwata-Gonokami}},\
  }\href@noop {} {\bibfield  {journal} {\bibinfo  {journal} {Nature
  Communications}\ }\textbf {\bibinfo {volume} {2}},\ \bibinfo {pages} {328}
  (\bibinfo {year} {2011})}\BibitemShut {NoStop}%
\bibitem [{\citenamefont {O'Hara}\ \emph {et~al.}(1999)\citenamefont {O'Hara},
  \citenamefont {Gullingsrud},\ and\ \citenamefont {Wolfe}}]{o1999auger}%
  \BibitemOpen
  \bibfield  {author} {\bibinfo {author} {\bibfnamefont {K.}~\bibnamefont
  {O'Hara}}, \bibinfo {author} {\bibfnamefont {J.}~\bibnamefont {Gullingsrud}},
  \ and\ \bibinfo {author} {\bibfnamefont {J.}~\bibnamefont {Wolfe}},\
  }\href@noop {} {\bibfield  {journal} {\bibinfo  {journal} {Physical Review
  B}\ }\textbf {\bibinfo {volume} {60}},\ \bibinfo {pages} {10872} (\bibinfo
  {year} {1999})}\BibitemShut {NoStop}%
\bibitem [{\citenamefont {Roslyak}\ \emph {et~al.}(2012)\citenamefont
  {Roslyak}, \citenamefont {Aparajita}, \citenamefont {Birman},\ and\
  \citenamefont {Mukamel}}]{roslyak2012coherent}%
  \BibitemOpen
  \bibfield  {author} {\bibinfo {author} {\bibfnamefont {O.}~\bibnamefont
  {Roslyak}}, \bibinfo {author} {\bibfnamefont {U.}~\bibnamefont {Aparajita}},
  \bibinfo {author} {\bibfnamefont {J.~L.}\ \bibnamefont {Birman}}, \ and\
  \bibinfo {author} {\bibfnamefont {S.}~\bibnamefont {Mukamel}},\ }\href@noop
  {} {\bibfield  {journal} {\bibinfo  {journal} {physica status solidi (b)}\
  }\textbf {\bibinfo {volume} {249}},\ \bibinfo {pages} {435} (\bibinfo {year}
  {2012})}\BibitemShut {NoStop}%
\bibitem [{\citenamefont {Naka}\ \emph {et~al.}(2013)\citenamefont {Naka},
  \citenamefont {Akimoto},\ and\ \citenamefont {Shirai}}]{naka2013free}%
  \BibitemOpen
  \bibfield  {author} {\bibinfo {author} {\bibfnamefont {N.}~\bibnamefont
  {Naka}}, \bibinfo {author} {\bibfnamefont {I.}~\bibnamefont {Akimoto}}, \
  and\ \bibinfo {author} {\bibfnamefont {M.}~\bibnamefont {Shirai}},\
  }\href@noop {} {\bibfield  {journal} {\bibinfo  {journal} {physica status
  solidi (b)}\ }\textbf {\bibinfo {volume} {250}},\ \bibinfo {pages} {1773}
  (\bibinfo {year} {2013})}\BibitemShut {NoStop}%
\bibitem [{\citenamefont {Mani}\ \emph {et~al.}(2010)\citenamefont {Mani},
  \citenamefont {Jang},\ and\ \citenamefont {Ketterson}}]{mani2010nonlinear}%
  \BibitemOpen
  \bibfield  {author} {\bibinfo {author} {\bibfnamefont {S.}~\bibnamefont
  {Mani}}, \bibinfo {author} {\bibfnamefont {J.}~\bibnamefont {Jang}}, \ and\
  \bibinfo {author} {\bibfnamefont {J.}~\bibnamefont {Ketterson}},\ }\href@noop
  {} {\bibfield  {journal} {\bibinfo  {journal} {Physical Review B}\ }\textbf
  {\bibinfo {volume} {82}},\ \bibinfo {pages} {113203} (\bibinfo {year}
  {2010})}\BibitemShut {NoStop}%
\bibitem [{\citenamefont {Goto}\ \emph {et~al.}(1997)\citenamefont {Goto},
  \citenamefont {Shen}, \citenamefont {Koyama},\ and\ \citenamefont
  {Yokouchi}}]{goto1997bose}%
  \BibitemOpen
  \bibfield  {author} {\bibinfo {author} {\bibfnamefont {T.}~\bibnamefont
  {Goto}}, \bibinfo {author} {\bibfnamefont {M.}~\bibnamefont {Shen}}, \bibinfo
  {author} {\bibfnamefont {S.}~\bibnamefont {Koyama}}, \ and\ \bibinfo {author}
  {\bibfnamefont {T.}~\bibnamefont {Yokouchi}},\ }\href@noop {} {\bibfield
  {journal} {\bibinfo  {journal} {Physical Review B}\ }\textbf {\bibinfo
  {volume} {55}},\ \bibinfo {pages} {7609} (\bibinfo {year}
  {1997})}\BibitemShut {NoStop}%
\bibitem [{\citenamefont {Fr{\"o}hlich}\ \emph {et~al.}(1991)\citenamefont
  {Fr{\"o}hlich}, \citenamefont {Kulik}, \citenamefont {Uebbing}, \citenamefont
  {Mysyrowicz}, \citenamefont {Langer}, \citenamefont {Stolz},\ and\
  \citenamefont {von~der Osten}}]{frohlich1991coherent}%
  \BibitemOpen
  \bibfield  {author} {\bibinfo {author} {\bibfnamefont {D.}~\bibnamefont
  {Fr{\"o}hlich}}, \bibinfo {author} {\bibfnamefont {A.}~\bibnamefont {Kulik}},
  \bibinfo {author} {\bibfnamefont {B.}~\bibnamefont {Uebbing}}, \bibinfo
  {author} {\bibfnamefont {A.}~\bibnamefont {Mysyrowicz}}, \bibinfo {author}
  {\bibfnamefont {V.}~\bibnamefont {Langer}}, \bibinfo {author} {\bibfnamefont
  {H.}~\bibnamefont {Stolz}}, \ and\ \bibinfo {author} {\bibfnamefont
  {W.}~\bibnamefont {von~der Osten}},\ }\href@noop {} {\bibfield  {journal}
  {\bibinfo  {journal} {Physical Review Letters}\ }\textbf {\bibinfo {volume}
  {67}},\ \bibinfo {pages} {2343} (\bibinfo {year} {1991})}\BibitemShut
  {NoStop}%
\bibitem [{\citenamefont {Courtens}\ \emph {et~al.}(1967)\citenamefont
  {Courtens}, \citenamefont {Bergman},\ and\ \citenamefont
  {Jortner}}]{courtens1967photo}%
  \BibitemOpen
  \bibfield  {author} {\bibinfo {author} {\bibfnamefont {E.}~\bibnamefont
  {Courtens}}, \bibinfo {author} {\bibfnamefont {A.}~\bibnamefont {Bergman}}, \
  and\ \bibinfo {author} {\bibfnamefont {J.}~\bibnamefont {Jortner}},\
  }\href@noop {} {\bibfield  {journal} {\bibinfo  {journal} {Physical Review}\
  }\textbf {\bibinfo {volume} {156}},\ \bibinfo {pages} {948} (\bibinfo {year}
  {1967})}\BibitemShut {NoStop}%
\bibitem [{\citenamefont {Kepler}(1967)}]{kepler1967photoionization}%
  \BibitemOpen
  \bibfield  {author} {\bibinfo {author} {\bibfnamefont {R.}~\bibnamefont
  {Kepler}},\ }\href@noop {} {\bibfield  {journal} {\bibinfo  {journal}
  {Physical Review Letters}\ }\textbf {\bibinfo {volume} {18}},\ \bibinfo
  {pages} {951} (\bibinfo {year} {1967})}\BibitemShut {NoStop}%
\bibitem [{\citenamefont {Schlotter}\ \emph {et~al.}(1977)\citenamefont
  {Schlotter}, \citenamefont {Kalinowski},\ and\ \citenamefont
  {B{\"a}ssler}}]{schlotter1977photoionization}%
  \BibitemOpen
  \bibfield  {author} {\bibinfo {author} {\bibfnamefont {P.}~\bibnamefont
  {Schlotter}}, \bibinfo {author} {\bibfnamefont {J.}~\bibnamefont
  {Kalinowski}}, \ and\ \bibinfo {author} {\bibfnamefont {H.}~\bibnamefont
  {B{\"a}ssler}},\ }\href@noop {} {\bibfield  {journal} {\bibinfo  {journal}
  {physica status solidi (b)}\ }\textbf {\bibinfo {volume} {81}},\ \bibinfo
  {pages} {521} (\bibinfo {year} {1977})}\BibitemShut {NoStop}%
\bibitem [{\citenamefont {Morikawa}\ \emph {et~al.}(1983)\citenamefont
  {Morikawa}, \citenamefont {Isono},\ and\ \citenamefont
  {Kotani}}]{morikawa1983generation}%
  \BibitemOpen
  \bibfield  {author} {\bibinfo {author} {\bibfnamefont {E.}~\bibnamefont
  {Morikawa}}, \bibinfo {author} {\bibfnamefont {Y.}~\bibnamefont {Isono}}, \
  and\ \bibinfo {author} {\bibfnamefont {M.}~\bibnamefont {Kotani}},\
  }\href@noop {} {\bibfield  {journal} {\bibinfo  {journal} {The Journal of
  chemical physics}\ }\textbf {\bibinfo {volume} {78}},\ \bibinfo {pages}
  {2691} (\bibinfo {year} {1983})}\BibitemShut {NoStop}%
\bibitem [{\citenamefont {Frazer}\ \emph {et~al.}(2014)\citenamefont {Frazer},
  \citenamefont {Schaller}, \citenamefont {Chang}, \citenamefont {Ketterson},\
  and\ \citenamefont {Poeppelmeier}}]{frazer2014third}%
  \BibitemOpen
  \bibfield  {author} {\bibinfo {author} {\bibfnamefont {L.}~\bibnamefont
  {Frazer}}, \bibinfo {author} {\bibfnamefont {R.~D.}\ \bibnamefont
  {Schaller}}, \bibinfo {author} {\bibfnamefont {K.~B.}\ \bibnamefont {Chang}},
  \bibinfo {author} {\bibfnamefont {J.~B.}\ \bibnamefont {Ketterson}}, \ and\
  \bibinfo {author} {\bibfnamefont {K.~R.}\ \bibnamefont {Poeppelmeier}},\
  }\href@noop {} {\bibfield  {journal} {\bibinfo  {journal} {Optics Letters}\
  }\textbf {\bibinfo {volume} {39}},\ \bibinfo {pages} {618} (\bibinfo {year}
  {2014})}\BibitemShut {NoStop}%
\bibitem [{\citenamefont {J{\"o}rger}\ \emph {et~al.}(2003)\citenamefont
  {J{\"o}rger}, \citenamefont {Tsitsishvili}, \citenamefont {Fleck},\ and\
  \citenamefont {Klingshirn}}]{jorger2003infrared}%
  \BibitemOpen
  \bibfield  {author} {\bibinfo {author} {\bibfnamefont {M.}~\bibnamefont
  {J{\"o}rger}}, \bibinfo {author} {\bibfnamefont {E.}~\bibnamefont
  {Tsitsishvili}}, \bibinfo {author} {\bibfnamefont {T.}~\bibnamefont {Fleck}},
  \ and\ \bibinfo {author} {\bibfnamefont {C.}~\bibnamefont {Klingshirn}},\
  }\href@noop {} {\bibfield  {journal} {\bibinfo  {journal} {physica status
  solidi (b)}\ }\textbf {\bibinfo {volume} {238}},\ \bibinfo {pages} {470}
  (\bibinfo {year} {2003})}\BibitemShut {NoStop}%
\bibitem [{\citenamefont {Kubouchi}\ \emph {et~al.}(2005)\citenamefont
  {Kubouchi}, \citenamefont {Yoshioka}, \citenamefont {Shimano}, \citenamefont
  {Mysyrowicz},\ and\ \citenamefont {Kuwata-Gonokami}}]{kubouchi2005study}%
  \BibitemOpen
  \bibfield  {author} {\bibinfo {author} {\bibfnamefont {M.}~\bibnamefont
  {Kubouchi}}, \bibinfo {author} {\bibfnamefont {K.}~\bibnamefont {Yoshioka}},
  \bibinfo {author} {\bibfnamefont {R.}~\bibnamefont {Shimano}}, \bibinfo
  {author} {\bibfnamefont {A.}~\bibnamefont {Mysyrowicz}}, \ and\ \bibinfo
  {author} {\bibfnamefont {M.}~\bibnamefont {Kuwata-Gonokami}},\ }\href@noop {}
  {\bibfield  {journal} {\bibinfo  {journal} {Physical Review Letters}\
  }\textbf {\bibinfo {volume} {94}},\ \bibinfo {pages} {016403} (\bibinfo
  {year} {2005})}\BibitemShut {NoStop}%
\bibitem [{\citenamefont {Kuwata-Gonokami}\ \emph {et~al.}(2004)\citenamefont
  {Kuwata-Gonokami}, \citenamefont {Kubouchi}, \citenamefont {Shimano},\ and\
  \citenamefont {Mysyrowicz}}]{kuwata2004time}%
  \BibitemOpen
  \bibfield  {author} {\bibinfo {author} {\bibfnamefont {M.}~\bibnamefont
  {Kuwata-Gonokami}}, \bibinfo {author} {\bibfnamefont {M.}~\bibnamefont
  {Kubouchi}}, \bibinfo {author} {\bibfnamefont {R.}~\bibnamefont {Shimano}}, \
  and\ \bibinfo {author} {\bibfnamefont {A.}~\bibnamefont {Mysyrowicz}},\
  }\href@noop {} {\bibfield  {journal} {\bibinfo  {journal} {Journal of the
  Physical Society of Japan}\ }\textbf {\bibinfo {volume} {73}},\ \bibinfo
  {pages} {1065} (\bibinfo {year} {2004})}\BibitemShut {NoStop}%
\bibitem [{\citenamefont {Tayagaki}\ \emph {et~al.}(2005)\citenamefont
  {Tayagaki}, \citenamefont {Mysyrowicz},\ and\ \citenamefont
  {Gonokami}}]{tayagaki2005yellow}%
  \BibitemOpen
  \bibfield  {author} {\bibinfo {author} {\bibfnamefont {T.}~\bibnamefont
  {Tayagaki}}, \bibinfo {author} {\bibfnamefont {A.}~\bibnamefont
  {Mysyrowicz}}, \ and\ \bibinfo {author} {\bibfnamefont {M.}~\bibnamefont
  {Gonokami}},\ }\href@noop {} {\bibfield  {journal} {\bibinfo  {journal}
  {Journal of the Physical Society of Japan}\ }\textbf {\bibinfo {volume}
  {74}},\ \bibinfo {pages} {1423} (\bibinfo {year} {2005})}\BibitemShut
  {NoStop}%
\bibitem [{\citenamefont {Yoshioka}\ and\ \citenamefont
  {Kuwata-Gonokami}(2006)}]{yoshioka2006dark}%
  \BibitemOpen
  \bibfield  {author} {\bibinfo {author} {\bibfnamefont {K.}~\bibnamefont
  {Yoshioka}}\ and\ \bibinfo {author} {\bibfnamefont {M.}~\bibnamefont
  {Kuwata-Gonokami}},\ }\href@noop {} {\bibfield  {journal} {\bibinfo
  {journal} {Physical Review B}\ }\textbf {\bibinfo {volume} {73}},\ \bibinfo
  {pages} {081202} (\bibinfo {year} {2006})}\BibitemShut {NoStop}%
\bibitem [{\citenamefont {Elliott}(1961)}]{elliott1961symmetry}%
  \BibitemOpen
  \bibfield  {author} {\bibinfo {author} {\bibfnamefont {R.}~\bibnamefont
  {Elliott}},\ }\href@noop {} {\bibfield  {journal} {\bibinfo  {journal}
  {Physical Review}\ }\textbf {\bibinfo {volume} {124}},\ \bibinfo {pages}
  {340} (\bibinfo {year} {1961})}\BibitemShut {NoStop}%
\bibitem [{\citenamefont {Landau}\ and\ \citenamefont
  {Lifshitz}(1977)}]{landau1977quantum}%
  \BibitemOpen
  \bibfield  {author} {\bibinfo {author} {\bibfnamefont {L.}~\bibnamefont
  {Landau}}\ and\ \bibinfo {author} {\bibfnamefont {E.}~\bibnamefont
  {Lifshitz}},\ }\href@noop {} {\emph {\bibinfo {title} {Quantum Mechanics,
  Section 142}}}\ (\bibinfo  {publisher} {Pergamon},\ \bibinfo {year}
  {1977})\BibitemShut {NoStop}%
\bibitem [{\citenamefont {Schmutzler}\ \emph {et~al.}(2013)\citenamefont
  {Schmutzler}, \citenamefont {Fr{\"o}hlich},\ and\ \citenamefont
  {Bayer}}]{schmutzler2013signatures}%
  \BibitemOpen
  \bibfield  {author} {\bibinfo {author} {\bibfnamefont {J.}~\bibnamefont
  {Schmutzler}}, \bibinfo {author} {\bibfnamefont {D.}~\bibnamefont
  {Fr{\"o}hlich}}, \ and\ \bibinfo {author} {\bibfnamefont {M.}~\bibnamefont
  {Bayer}},\ }\href@noop {} {\bibfield  {journal} {\bibinfo  {journal}
  {Physical Review B}\ }\textbf {\bibinfo {volume} {87}},\ \bibinfo {pages}
  {245202} (\bibinfo {year} {2013})}\BibitemShut {NoStop}%
\bibitem [{\citenamefont {Combescot}\ \emph {et~al.}(2007)\citenamefont
  {Combescot}, \citenamefont {Dupertuis},\ and\ \citenamefont
  {Betbeder-Matibet}}]{combescot2007polariton}%
  \BibitemOpen
  \bibfield  {author} {\bibinfo {author} {\bibfnamefont {M.}~\bibnamefont
  {Combescot}}, \bibinfo {author} {\bibfnamefont {M.}~\bibnamefont
  {Dupertuis}}, \ and\ \bibinfo {author} {\bibfnamefont {O.}~\bibnamefont
  {Betbeder-Matibet}},\ }\href@noop {} {\bibfield  {journal} {\bibinfo
  {journal} {EPL}\ }\textbf {\bibinfo {volume} {79}},\ \bibinfo {pages} {17001}
  (\bibinfo {year} {2007})}\BibitemShut {NoStop}%
\bibitem [{\citenamefont {Moskalenko}\ and\ \citenamefont
  {Snoke}(2000)}]{moskalenko2000bose}%
  \BibitemOpen
  \bibfield  {author} {\bibinfo {author} {\bibfnamefont {S.~A.}\ \bibnamefont
  {Moskalenko}}\ and\ \bibinfo {author} {\bibfnamefont {D.~W.}\ \bibnamefont
  {Snoke}},\ }\href@noop {} {\emph {\bibinfo {title} {Bose-Einstein
  Condensation of Excitons and Biexcitons and Coherent Nonlinear Optics with
  Excitons}}}\ (\bibinfo  {publisher} {Cambridge University Press},\ \bibinfo
  {year} {2000})\BibitemShut {NoStop}%
\bibitem [{\citenamefont {Kuwata-Gonokami}\ \emph {et~al.}(2002)\citenamefont
  {Kuwata-Gonokami}, \citenamefont {Shimano},\ and\ \citenamefont
  {Mysyrowicz}}]{kuwata2002phase}%
  \BibitemOpen
  \bibfield  {author} {\bibinfo {author} {\bibfnamefont {M.}~\bibnamefont
  {Kuwata-Gonokami}}, \bibinfo {author} {\bibfnamefont {R.}~\bibnamefont
  {Shimano}}, \ and\ \bibinfo {author} {\bibfnamefont {A.}~\bibnamefont
  {Mysyrowicz}},\ }\href@noop {} {\bibfield  {journal} {\bibinfo  {journal}
  {Journal of the Physical Society of Japan}\ }\textbf {\bibinfo {volume}
  {71}},\ \bibinfo {pages} {1257} (\bibinfo {year} {2002})}\BibitemShut
  {NoStop}%
\bibitem [{\citenamefont {Yoshioka}\ \emph {et~al.}(2014)\citenamefont
  {Yoshioka}, \citenamefont {Miyashita},\ and\ \citenamefont
  {Kuwata-Gonokami}}]{yoshioka2014selective}%
  \BibitemOpen
  \bibfield  {author} {\bibinfo {author} {\bibfnamefont {K.}~\bibnamefont
  {Yoshioka}}, \bibinfo {author} {\bibfnamefont {K.}~\bibnamefont {Miyashita}},
  \ and\ \bibinfo {author} {\bibfnamefont {M.}~\bibnamefont
  {Kuwata-Gonokami}},\ }\href@noop {} {\bibfield  {journal} {\bibinfo
  {journal} {Optics Express}\ }\textbf {\bibinfo {volume} {22}},\ \bibinfo
  {pages} {3261} (\bibinfo {year} {2014})}\BibitemShut {NoStop}%
\bibitem [{\citenamefont {Halasyamani}\ and\ \citenamefont
  {Poeppelmeier}(1998)}]{halasyamani1998noncentrosymmetric}%
  \BibitemOpen
  \bibfield  {author} {\bibinfo {author} {\bibfnamefont {P.~S.}\ \bibnamefont
  {Halasyamani}}\ and\ \bibinfo {author} {\bibfnamefont {K.~R.}\ \bibnamefont
  {Poeppelmeier}},\ }\href@noop {} {\bibfield  {journal} {\bibinfo  {journal}
  {Chemistry of Materials}\ }\textbf {\bibinfo {volume} {10}},\ \bibinfo
  {pages} {2753} (\bibinfo {year} {1998})}\BibitemShut {NoStop}%
\bibitem [{\citenamefont {Epperlein}\ \emph {et~al.}(1987)\citenamefont
  {Epperlein}, \citenamefont {Dick}, \citenamefont {Marowsky},\ and\
  \citenamefont {Reider}}]{epperlein1987second}%
  \BibitemOpen
  \bibfield  {author} {\bibinfo {author} {\bibfnamefont {D.}~\bibnamefont
  {Epperlein}}, \bibinfo {author} {\bibfnamefont {B.}~\bibnamefont {Dick}},
  \bibinfo {author} {\bibfnamefont {G.}~\bibnamefont {Marowsky}}, \ and\
  \bibinfo {author} {\bibfnamefont {G.}~\bibnamefont {Reider}},\ }\href@noop {}
  {\bibfield  {journal} {\bibinfo  {journal} {Applied Physics B}\ }\textbf
  {\bibinfo {volume} {44}},\ \bibinfo {pages} {5} (\bibinfo {year}
  {1987})}\BibitemShut {NoStop}%
\bibitem [{\citenamefont {Fr{\"o}hlich}\ \emph {et~al.}(2005)\citenamefont
  {Fr{\"o}hlich}, \citenamefont {Dasbach}, \citenamefont {Baldassarri
  H{\"o}ger~von H{\"o}gersthal}, \citenamefont {Bayer}, \citenamefont
  {Klieber}, \citenamefont {Suter},\ and\ \citenamefont
  {Stolz}}]{frohlich2005high}%
  \BibitemOpen
  \bibfield  {author} {\bibinfo {author} {\bibfnamefont {D.}~\bibnamefont
  {Fr{\"o}hlich}}, \bibinfo {author} {\bibfnamefont {G.}~\bibnamefont
  {Dasbach}}, \bibinfo {author} {\bibfnamefont {G.}~\bibnamefont {Baldassarri
  H{\"o}ger~von H{\"o}gersthal}}, \bibinfo {author} {\bibfnamefont
  {M.}~\bibnamefont {Bayer}}, \bibinfo {author} {\bibfnamefont
  {R.}~\bibnamefont {Klieber}}, \bibinfo {author} {\bibfnamefont
  {D.}~\bibnamefont {Suter}}, \ and\ \bibinfo {author} {\bibfnamefont
  {H.}~\bibnamefont {Stolz}},\ }\href@noop {} {\bibfield  {journal} {\bibinfo
  {journal} {Solid State Communications}\ }\textbf {\bibinfo {volume} {134}},\
  \bibinfo {pages} {139} (\bibinfo {year} {2005})}\BibitemShut {NoStop}%
\bibitem [{\citenamefont {Chang}\ \emph {et~al.}(2013)\citenamefont {Chang},
  \citenamefont {Frazer}, \citenamefont {Schwartz}, \citenamefont {Ketterson},\
  and\ \citenamefont {Poeppelmeier}}]{chang2013removal}%
  \BibitemOpen
  \bibfield  {author} {\bibinfo {author} {\bibfnamefont {K.~B.}\ \bibnamefont
  {Chang}}, \bibinfo {author} {\bibfnamefont {L.}~\bibnamefont {Frazer}},
  \bibinfo {author} {\bibfnamefont {J.~J.}\ \bibnamefont {Schwartz}}, \bibinfo
  {author} {\bibfnamefont {J.~B.}\ \bibnamefont {Ketterson}}, \ and\ \bibinfo
  {author} {\bibfnamefont {K.~R.}\ \bibnamefont {Poeppelmeier}},\ }\href@noop
  {} {\bibfield  {journal} {\bibinfo  {journal} {Crystal Growth \& Design}\
  }\textbf {\bibinfo {volume} {13}},\ \bibinfo {pages} {4914} (\bibinfo {year}
  {2013})}\BibitemShut {NoStop}%
\bibitem [{\citenamefont {Ito}\ and\ \citenamefont
  {Masumi}(1997)}]{ito1997detailed}%
  \BibitemOpen
  \bibfield  {author} {\bibinfo {author} {\bibfnamefont {T.}~\bibnamefont
  {Ito}}\ and\ \bibinfo {author} {\bibfnamefont {T.}~\bibnamefont {Masumi}},\
  }\href@noop {} {\bibfield  {journal} {\bibinfo  {journal} {Journal of the
  Physical Society of Japan}\ }\textbf {\bibinfo {volume} {66}},\ \bibinfo
  {pages} {2185} (\bibinfo {year} {1997})}\BibitemShut {NoStop}%
\bibitem [{\citenamefont {Kramida}\ \emph {et~al.}(2013)\citenamefont
  {Kramida}, \citenamefont {{Yu. Ralchenko}}, \citenamefont {Reader},\ and\
  \citenamefont {{NIST ASD Team}}}]{NIST_ASD}%
  \BibitemOpen
  \bibfield  {author} {\bibinfo {author} {\bibfnamefont {A.}~\bibnamefont
  {Kramida}}, \bibinfo {author} {\bibnamefont {{Yu. Ralchenko}}}, \bibinfo
  {author} {\bibfnamefont {J.}~\bibnamefont {Reader}}, \ and\ \bibinfo {author}
  {\bibnamefont {{NIST ASD Team}}},\ }\href@noop {} {}\bibinfo {howpublished}
  {{NIST Atomic Spectra Database (ver. 5.1), [Online]. Available:
  {\tt{http://physics.nist.gov/asd}} [2013, December 22]. National Institute of
  Standards and Technology, Gaithersburg, MD.}} (\bibinfo {year}
  {2013})\BibitemShut {NoStop}%
\bibitem [{\citenamefont {Rakhshani}(1986)}]{rakhshani1986preparation}%
  \BibitemOpen
  \bibfield  {author} {\bibinfo {author} {\bibfnamefont {A.}~\bibnamefont
  {Rakhshani}},\ }\href@noop {} {\bibfield  {journal} {\bibinfo  {journal}
  {Solid-State Electronics}\ }\textbf {\bibinfo {volume} {29}},\ \bibinfo
  {pages} {7} (\bibinfo {year} {1986})}\BibitemShut {NoStop}%
\bibitem [{\citenamefont {Wei}\ \emph {et~al.}(2012)\citenamefont {Wei},
  \citenamefont {Gong}, \citenamefont {Chen}, \citenamefont {Zi},\ and\
  \citenamefont {Cao}}]{wei2012photovoltaic}%
  \BibitemOpen
  \bibfield  {author} {\bibinfo {author} {\bibfnamefont {H.}~\bibnamefont
  {Wei}}, \bibinfo {author} {\bibfnamefont {H.}~\bibnamefont {Gong}}, \bibinfo
  {author} {\bibfnamefont {L.}~\bibnamefont {Chen}}, \bibinfo {author}
  {\bibfnamefont {M.}~\bibnamefont {Zi}}, \ and\ \bibinfo {author}
  {\bibfnamefont {B.}~\bibnamefont {Cao}},\ }\href@noop {} {\bibfield
  {journal} {\bibinfo  {journal} {The Journal of Physical Chemistry C}\
  }\textbf {\bibinfo {volume} {116}},\ \bibinfo {pages} {10510} (\bibinfo
  {year} {2012})}\BibitemShut {NoStop}%
\bibitem [{\citenamefont {Septina}\ \emph {et~al.}(2011)\citenamefont
  {Septina}, \citenamefont {Ikeda}, \citenamefont {Khan}, \citenamefont
  {Hirai}, \citenamefont {Harada}, \citenamefont {Matsumura},\ and\
  \citenamefont {Peter}}]{septina2011potentiostatic}%
  \BibitemOpen
  \bibfield  {author} {\bibinfo {author} {\bibfnamefont {W.}~\bibnamefont
  {Septina}}, \bibinfo {author} {\bibfnamefont {S.}~\bibnamefont {Ikeda}},
  \bibinfo {author} {\bibfnamefont {M.~A.}\ \bibnamefont {Khan}}, \bibinfo
  {author} {\bibfnamefont {T.}~\bibnamefont {Hirai}}, \bibinfo {author}
  {\bibfnamefont {T.}~\bibnamefont {Harada}}, \bibinfo {author} {\bibfnamefont
  {M.}~\bibnamefont {Matsumura}}, \ and\ \bibinfo {author} {\bibfnamefont
  {L.~M.}\ \bibnamefont {Peter}},\ }\href@noop {} {\bibfield  {journal}
  {\bibinfo  {journal} {Electrochimica Acta}\ }\textbf {\bibinfo {volume}
  {56}},\ \bibinfo {pages} {4882} (\bibinfo {year} {2011})}\BibitemShut
  {NoStop}%
\bibitem [{\citenamefont {Lee}\ \emph {et~al.}(2012)\citenamefont {Lee},
  \citenamefont {Winkler}, \citenamefont {Cheng~Siah}, \citenamefont {Brandt},\
  and\ \citenamefont {Buonassisi}}]{lee2012growth}%
  \BibitemOpen
  \bibfield  {author} {\bibinfo {author} {\bibfnamefont {Y.~S.}\ \bibnamefont
  {Lee}}, \bibinfo {author} {\bibfnamefont {M.~T.}\ \bibnamefont {Winkler}},
  \bibinfo {author} {\bibfnamefont {S.}~\bibnamefont {Cheng~Siah}}, \bibinfo
  {author} {\bibfnamefont {R.}~\bibnamefont {Brandt}}, \ and\ \bibinfo {author}
  {\bibfnamefont {T.}~\bibnamefont {Buonassisi}},\ }in\ \href@noop {} {\emph
  {\bibinfo {booktitle} {Photovoltaic Specialists Conference (PVSC), 2012 38th
  IEEE}}}\ (\bibinfo {organization} {IEEE},\ \bibinfo {year} {2012})\ pp.\
  \bibinfo {pages} {002557--002558}\BibitemShut {NoStop}%
\bibitem [{\citenamefont {Bendavid}\ and\ \citenamefont
  {Carter}(2013)}]{bendavid2013first}%
  \BibitemOpen
  \bibfield  {author} {\bibinfo {author} {\bibfnamefont {L.~I.}\ \bibnamefont
  {Bendavid}}\ and\ \bibinfo {author} {\bibfnamefont {E.~A.}\ \bibnamefont
  {Carter}},\ }\href@noop {} {\bibfield  {journal} {\bibinfo  {journal} {The
  Journal of Physical Chemistry B}\ }\textbf {\bibinfo {volume} {117}},\
  \bibinfo {pages} {15750} (\bibinfo {year} {2013})}\BibitemShut {NoStop}%
\bibitem [{\citenamefont {Zhu}\ \emph {et~al.}(2012)\citenamefont {Zhu},
  \citenamefont {Zhang}, \citenamefont {Lv}, \citenamefont {Chu}, \citenamefont
  {Ye},\ and\ \citenamefont {Zhou}}]{zhu2012cuprous}%
  \BibitemOpen
  \bibfield  {author} {\bibinfo {author} {\bibfnamefont {Q.}~\bibnamefont
  {Zhu}}, \bibinfo {author} {\bibfnamefont {Y.}~\bibnamefont {Zhang}}, \bibinfo
  {author} {\bibfnamefont {F.}~\bibnamefont {Lv}}, \bibinfo {author}
  {\bibfnamefont {P.~K.}\ \bibnamefont {Chu}}, \bibinfo {author} {\bibfnamefont
  {Z.}~\bibnamefont {Ye}}, \ and\ \bibinfo {author} {\bibfnamefont
  {F.}~\bibnamefont {Zhou}},\ }\href@noop {} {\bibfield  {journal} {\bibinfo
  {journal} {Journal of Hazardous Materials}\ }\textbf {\bibinfo {volume}
  {217}},\ \bibinfo {pages} {11} (\bibinfo {year} {2012})}\BibitemShut
  {NoStop}%
\bibitem [{\citenamefont {Pal}\ \emph {et~al.}(2013)\citenamefont {Pal},
  \citenamefont {Ganguly}, \citenamefont {Mondal}, \citenamefont {Roy},
  \citenamefont {Negishi},\ and\ \citenamefont {Pal}}]{pal2013crystal}%
  \BibitemOpen
  \bibfield  {author} {\bibinfo {author} {\bibfnamefont {J.}~\bibnamefont
  {Pal}}, \bibinfo {author} {\bibfnamefont {M.}~\bibnamefont {Ganguly}},
  \bibinfo {author} {\bibfnamefont {C.}~\bibnamefont {Mondal}}, \bibinfo
  {author} {\bibfnamefont {A.}~\bibnamefont {Roy}}, \bibinfo {author}
  {\bibfnamefont {Y.}~\bibnamefont {Negishi}}, \ and\ \bibinfo {author}
  {\bibfnamefont {T.}~\bibnamefont {Pal}},\ }\href@noop {} {\bibfield
  {journal} {\bibinfo  {journal} {The Journal of Physical Chemistry C}\
  }\textbf {\bibinfo {volume} {117}},\ \bibinfo {pages} {24640} (\bibinfo
  {year} {2013})}\BibitemShut {NoStop}%
\bibitem [{\citenamefont {Yoon}\ \emph {et~al.}(2014)\citenamefont {Yoon},
  \citenamefont {Kim}, \citenamefont {Kim}, \citenamefont {Lim},\ and\
  \citenamefont {Yoo}}]{yoon2014manipulation}%
  \BibitemOpen
  \bibfield  {author} {\bibinfo {author} {\bibfnamefont {S.}~\bibnamefont
  {Yoon}}, \bibinfo {author} {\bibfnamefont {M.}~\bibnamefont {Kim}}, \bibinfo
  {author} {\bibfnamefont {I.-S.}\ \bibnamefont {Kim}}, \bibinfo {author}
  {\bibfnamefont {J.}~\bibnamefont {Lim}}, \ and\ \bibinfo {author}
  {\bibfnamefont {B.}~\bibnamefont {Yoo}},\ }\href@noop {} {\bibfield
  {journal} {\bibinfo  {journal} {Journal of Materials Chemistry A}\ }
  (\bibinfo {year} {2014})}\BibitemShut {NoStop}%
\end{thebibliography}%

\end{document}